\documentclass[conference]{IEEEtran}

\usepackage{longtable,booktabs}
\usepackage{graphicx}
\usepackage{comment}
\usepackage{xspace}
\usepackage{todonotes}
\usepackage{xifthen}
\usepackage{fancyvrb}
\usepackage{arrayjobx}
\usepackage{soul}
\usepackage{xcolor}
\usepackage{subfig}
\usepackage{flushend}
\usepackage{placeins}
\usepackage{ifthen}

\newcommand{\mate}{MATE\xspace}
\newcommand{\etal}{{\em et al.}\xspace}
\newcommand{\cb}{Code\hspace{0pt}::\hspace{0pt}Blocks\xspace}

\makeatletter

\let\oldcap\caption

\def\longtable[#1]#2{%
	\begin{table}[tb]
		\centering
		\def\caption##1{%
			\oldcap{##1}
			\begin{tabular}{#2}
			}
		}
		\def\endlongtable{\end{tabular}\end{table}}
\makeatother

\newcommand{\var}{Var\-Merge\xspace}
\newcommand{\perceived}{perceived\xspace}

\newarray\tasksub
\readarray{tasksub}{C,Lo&O,Lo&C,Ly&O,Ly}
\newcommand{\task}[1][]{
	\ifthenelse{\isempty{#1}}
	{\mathcal{T}}
	{\mathcal{T}_{\tasksub(#1)}}
}

\usepackage{amsmath}
\usepackage{algorithmic}
\usepackage{times}

\newboolean{visualizzamodifiche}
\newcommand{\MODIFICATO}[1]{\ifthenelse{\boolean{visualizzamodifiche}}{\textcolor{blue}{#1}}{#1}}

\setboolean{visualizzamodifiche}{false}

\title{Assessment of Source Code Obfuscation Techniques}

\author{%
	\IEEEauthorblockN{%
		Alessio Viticchi\'e\IEEEauthorrefmark{1},
		Leonardo Regano\IEEEauthorrefmark{1},
		Marco Torchiano\IEEEauthorrefmark{1},
		Cataldo Basile\IEEEauthorrefmark{1},\\
		Mariano Ceccato\IEEEauthorrefmark{2},
		Paolo Tonella\IEEEauthorrefmark{2}
		and
		Roberto Tiella\IEEEauthorrefmark{2}\\
		~
	}
	\IEEEauthorblockA{%
		\IEEEauthorrefmark{1}%
		Dipartimento di Automatica e Informatica, \\
		Politecnico di Torino, Torino, Italy   \\
		\{first.last\}@polito.it \\~
	}
	\IEEEauthorblockA{%
		\IEEEauthorrefmark{2}%
		FBK, Trento, Italy \\
		\{last\}@fbk.eu
	}

}

\begin{document}


\maketitle

\begin{abstract}
%
%
%
%
%
Obfuscation techniques are a general category of software protections widely adopted to prevent malicious tampering of the code by making applications more difficult to understand and thus harder to modify.
Obfuscation techniques are divided in code and data obfuscation, depending on the protected asset.
While preliminary empirical studies have been conducted to determine the impact of code obfuscation, our work aims at assessing the effectiveness and efficiency in preventing attacks of a specific data obfuscation technique -- \var.
We conducted an experiment with student participants performing two attack tasks on clear and obfuscated versions of two applications written in C.
The experiment showed a significant effect of data obfuscation on both the time required to complete and the successful attack efficiency.
An application with \var reduces by six times the number of successful attacks per unit of time.
This outcome provides a practical clue that can be used when applying software protections based on data obfuscation.
\end{abstract}


%
\IEEEpeerreviewmaketitle

\section{Introduction}
\label{sec:intro}

Recently, a new class of attacks against software has received great attention: Man-At-The-End attacks (\mate) \cite{mate-attacks,falcarin2011guest}.
\mate attacks aim at compromising software assets, which can be classified in two types: data and code.
Software developers aim at protecting assets' security properties, namely data confidentiality, privacy, integrity, and code confidentiality, execution correctness, and integrity.
\mate attacks are more powerful than the ones against traditional cryptography, the Man-In-The-Middle attacks, as 
\mate attackers are the end users. Hence, they have full privileges and full access to executables on their platforms, where they also have at their disposal a vast set of tools, like debuggers, decompilers, and static and dynamic analysers.

Obfuscation is one of the most used protection technique to prevent the comprehension of programs against \mate attacks.
There are several approaches to obfuscation \cite{Collberg2009Surreptitious}, which can be mainly divided in code obfuscation and data obfuscation.
All obfuscation types transform a program in such a way that it is more difficult to understand for an attacker, while at the same time preserving its original program functionality.
However, data obfuscation transformations change programs with the aim of hiding both variable content and usage, while code obfuscation changes layout and control information to render the code difficult to reverse engineer and understand.
The hypothesis is that obfuscation can protect asset confidentiality by making it difficult for attackers to understand the application under attack.
Obfuscation is also used indirectly to protect data and code integrity, as well as execution correctness, since changing programs' functionality is also difficult when the code is difficult to understand.
Ultimately, obfuscation discourages attacks against software assets as it renders them less economically favourable.
Indeed, mounting successful attacks against software assets protected with obfuscation requires more sophisticated attack tools, and more time to comprehend the assets, thus a greater investment. As a consequence, profits from exploiting software attacks are reduced.


Obfuscation makes attacks \emph{more} complex but it cannot completely block them \cite{barak2001pop}.
Researchers have focused their attention on how much obfuscation is effective in protecting software assets.
Initially, the impact of obfuscation has been extensively investigated by means of traditional software assessment techniques, based on internal code metrics~\cite{collberg1997taxonomy,anckaert2007obfuscation,linn2003obfuscation,goto2000quantitative,udupa2005deobfuscation,visaggio2013empirical,ceccato2014large}.
The claim is that by means of these methods an objective evaluation of tamper resistance can be reached.
However, these types of quantification of software protections are unable to entirely capture the actual complexity for attackers who aim at compromising the assets.
A program with higher code complexity (e.g., one protected with control flow obfuscation) should require at least the same time to mount attacks, but there is no evidence of significant correlation between complexity metrics and actual attack delays.
In other words, impact of code protections metrics was not empirically validated.

Later, empirical studies have been conducted~\cite{sutherland2006empirical,ceccato2009effectiveness,CeccatoPFRTT14} to assess the impact of code obfuscation on delaying the completion of attack tasks.
These studies have targeted a quantitative estimation of the cost of understanding obfuscated code, by means of controlled experiments with human subjects.
Human subjects have been asked to perform tasks on ad hoc defined and protected applications. Then, the impact of code obfuscation has been estimated by means of two main parameters: \textit{correctness}, which counts the number of successful attacks, and \textit{efficiency}, which measures the delay added by the presence of obfuscation.


This paper assesses the effectiveness and efficiency of the \var data obfuscation technique by comparing the time needed to mount attack tasks on clear and obfuscated versions of two applications written in C, and assessing the success rate in the execution of the task.
The \var technique has been selected among a set of candidate techniques as one of the most effective ones (using Collberg's terminology, with high potency) and is applicable to C source code (a prerequisite for the involvement of students as subjects at our institutions).
While it is a general feeling that obfuscation renders programs more difficult to understand, we wanted to confirm this hypothesis on a data obfuscation technique and derive, if possible, quantitative measures of its impact.
To the best of our knowledge, this is the first work that aims at estimating with an empirical study the effectiveness of a data obfuscation technique. While previous work focused on control-flow obfuscation.

The paper is organised as follows.
Section~\ref{sec:background} presents previous works on the assessment of obfuscation effectiveness. 
Section~\ref{sec:design} details the experiments preparation, realisation, and analysis method.
In Section~\ref{results}, the results of the experiments are analysed to confirm or reject the research hypothesis.
Section~\ref{sec:discussion} discusses the results and the limitations that might affect their validity.
Finally, Section~\ref{sec:conclusions} draws conclusions and presents future experiments that may lead to a more comprehensive assessment of obfuscation techniques.

%
\section{Background}
\label{sec:background}

\subsection{Related work}

There are two main research approaches for the assessment of obfuscation techniques: assessment based on internal software metrics  and assessment with experiments involving human subjects.

The first approach has been introduced by Collberg \etal, who defined the concept of {\em potency}, a metric to estimate the effectiveness of obfuscation on programs \cite{collberg1997taxonomy}.
Potency has been later used by Anckaert \etal to compare obfuscation techniques \cite{anckaert2007obfuscation}.
Linn \etal  considered the confusion factor, which estimates the number of binary instructions that a code decompiler is not able to parse \cite{linn2003obfuscation}.
Goto \etal  proposed a method to quantitatively measure the complexity of obfuscated code based on the compiler syntax analysis \cite{goto2000quantitative}.
Udupa \etal estimated the increase of complexity in obfuscated programs by using data that can be extracted with static and dynamic analysis tools \cite{udupa2005deobfuscation}.
Visaggio \etal instead used  code entropy as a protection potency metric for obfuscated Javascript code \cite{visaggio2013empirical}.
Ceccato \etal  evaluated the impact of several obfuscation techniques on Java code quality \cite{ceccato2014large}. The authors performed a large set of experiments and estimated the effects of obfuscation on ten different complexity and modularity metrics.
%

Assessment by means of experiments with human subjects has been first presented in a work by Sutherland \etal, who published the first study with human subjects~\cite{sutherland2006empirical}. The authors correlated the expertise of attackers with the correctness of reverse engineering tasks. Moreover, they proved that source code metrics are not appropriate to estimate the delays on attack tasks, when binary code is involved.
Ceccato \etal measured, with two controlled experiments, the correctness and effectiveness in understanding and modifying decompiled obfuscated Java code, compared to decompiled clear code \cite{ceccato2009effectiveness}.
This work has been extended with a larger set of experiments on several obfuscation techniques in a successive work \cite{CeccatoPFRTT14}.
The major difference with respect to previous work is the obfuscation used in the study.
In fact, Ceccato \etal studied obfuscations aiming at hiding control flow and variable names. Conversely, we focus on a transformation meant to hide values of critical variables.
\MODIFICATO{Our work continues the effort in empirically assessing the effectiveness of protection techniques by means of experiments involving human subjects, by addressing a technique, \var, which was not assessed before, in a category of obfuscation techniques, data obfuscation, which was not yet target of experiments.}


\subsection{Obfuscation}\label{sec:obfuscation}

\begin{figure*}[t]
\centering
\begin{minipage}{0.3\linewidth}
\begin{Verbatim}[frame=single,numbers=left,numbersep=2pt,label=clear.c]
int main(){
  int a,b;

  a = 3;
  b = 5;

  printf("%d\n",a+b);
}
\end{Verbatim}
\end{minipage}
\quad
\begin{minipage}{0.6\linewidth}
\begin{Verbatim}[frame=single,numbers=left,numbersep=2pt,label=obfuscated.c]
int main(){
  unsigned long x;

  /*  3L | (5L << 0x20) */
  x = 21474836483L;

  printf("%d\n",(int)(x&0xffff)+(int)(x>>0x20));
}
\end{Verbatim}
\end{minipage}
\caption{Example of clear and obfuscated code using \var}
\label{fig:samplecode}
\end{figure*}

As anticipated, data obfuscation aims at hiding both variable content and usage.
Data obfuscation can be applied for instance to critical data, such as: user-IDs, counters, expiration dates, or privacy-sensitive data such as medical data.
On the other hand, data obfuscation is not suited for hiding cryptographic keys, as in this case white box cryptography offers a stronger protection for this specific asset.
Several data obfuscation transformations have been proposed in the  literature. They have been initially classified as \cite{collberg1997taxonomy}:
\begin{itemize}
 \item \textbf{Storage \& Encoding}: change representation of (sca\-lar) data;
 \item \textbf{Aggregation}: alter how data (both scalar variables and arrays) are aggregated;
 \item \textbf{Ordering}: permute items in existing data structures. 
\end{itemize}

In this paper, we report the empirical validation of the effectiveness of a specific aggregation technique, namely \var, which merges several scalar variables into a single one.
When selecting this technique, we have only considered Storage \& Encoding and Aggregation techniques that are not targeted for the Object Oriented paradigm.
We have excluded Reordering transformations as their potency is low.
The candidate techniques were: `Split variables', `Change encoding', `Change variable lifetimes', `Convert static data to procedure', \var, and `Split, fold, merge, arrays'.
All the advantages and disadvantages of these techniques have been considered and, in the end, \var has been selected because of its high potency, the reasonable effort to implement an automatic data obfuscator, and because it can be applied to the source code of applications written in C, the language mastered by students at our institutions.
All these decisions have been made and validated within the context of the ASPIRE project\footnote{www.aspire-fp7.eu}.

Given $n$ variables $v_1,...,v_n$ whose domain is represented by $b(v_1)$,...,$b(v_n)$ bits, \var creates a new variable $m$, whose domain is $b(m)=\sum_i^n b(v_1)$,
\var performs the following aggregation:
$$m=v_1+v_2\cdot 2^{b(v_1)}+...+v_n\cdot 2^{\sum_i^{n-1} b(v_1)}$$
Operations on the original variables can be mapped to operations on the merging variable by means of proper mask and shift operations.

Figure \ref{fig:samplecode} reports an example of clear code and the corresponding obfuscated code. We can observe how the original variables as well as their values is no more clearly available.

The effectiveness of \var relies on the fact that it breaks the easy association between a variable and its use, that is, its semantics.
Since most of the operations in the obfuscated code involve the use of the same variables with different shifts, an attacker is obliged to understand the code semantics by reconstructing the variable flows and by associating values with shifts. Moreover, proper bitwise operations are needed to obtain the actual decimal values.
Of course, expert attackers may recognise that a program has been protected with \var by the unusual number of shift operations. Correspondingly, they may adapt their attack strategy to the recognised protection.
Nevertheless, they have to invest non-negligible time to analyse the aggregation variables and to determine the shift offsets necessary to reconstruct the program semantics.




\section{Experiment Design}
\label{sec:design}
The next sections present all the preparation and realisation phases of the experiment.

%
%

%
%
%
%
\subsection{Goal and Research Questions}
The main \emph{goal} of the study is to evaluate the effect of a specific source code obfuscation technique,
\var, with the \emph{purpose} of evaluating its ability of making the code resilient
to malicious attacks.
The \emph{quality focus} is  how the technique reduces the attacker's capability to
successfully perform an attack by forcing the application behaviour.


The study is interpreted from the \emph{perspective} of an attacker, since we aim at evaluating the increased difficulty \perceived by attackers
when the code is protected by \var.
In particular, in our case the role of the attacker is played by a set of students that have a consolidated
minimum level of expertise in manipulating application source code.

\subsubsection{The subjects}
The \emph{context} of the study consists of \emph{subjects}, i.e., the students acting as attackers, who perform their attacks on  \emph{objects}, i.e., the systems to be attacked.

Subjects are 15 University students: 14 Master students in Computer Science Engineering and 1 PhD student in Computer and Control Engineering,
both from  Politecnico di Torino.
\MODIFICATO{All the students are knowledgeable about C programming and software engineering. We filtered Master and PhD students based on their academic career and grades.
Supported by the fact that  Politecnico di Torino offers and requires  a strong knowledge of programming, in particular  C programming, in many courses, we estimated their expertise as sufficient to perform  the tasks that we were proposing them.
Moreover, we precisely estimated their expertise during the results assessment phase based on their expertise self estimation and number of years of experience, which is the best practice according to a previous work \cite{feigenspan2012measuring}.}

The subjects are not expected to have any knowledge about \mate scenarios, attacks, and attacks strategies.
Indeed, the students of Politecnico di Torino do not attend any course about software tampering or software reverse engineering.
Thus, students are probably not the best choice to model real subjects.
Professional hackers could be better subjects to evaluate \mate attacks exploitation, but it is considerably difficult to involve them.
We considered the competences and capabilities of our subjects during the design and the analysis, in particular when selecting the applications and the tasks. We think the use of students as subjects did not affect our main conclusions, as explained in
Section~\ref{sec:threatstovalidity}, since we measure the impact of \var on attack time in a comparative way.

We encouraged the students' participation by putting up for grab two gift certificates among all the participants regardless of the success in performing their attack tasks.
In our opinion, the prize encourages  participation and, at the same time, the possibility of a win induces a larger commitment in the tasks.
On the other hand, assigning prizes to all participants would lead to higher participation, but it would introduce noise into the collected data,
as subjects might participate just for prize.
Finally, giving no incentives at all was not possible, as students do not usually like to spend extra time on non-profitable academic tasks instead of investing it in regular academic activities.
In conclusion, university students participation
had to be stimulated in some way, since they are students and not professional
attackers (such as Tiger teams, penetration testers or ethic hackers)
as they do not get any monetary advantages in performing the experiment tasks.

\subsubsection{The objects}
\label{sec:objects}

The \emph{objects} of the experiment are two applications written in C.
We define as \emph{clear} the original application with no obfuscation applied and as \emph{obfuscated}
the application on which \var is applied.

The first application, named \emph{Lotto}, is a stand-alone lottery game.
This game allows the user to input a sequence of seven numbers (six numbers plus one bonus number) and tells, as output, if the sequence matches another sequence, named the jackpot sequence, which is hard-coded in an array variable inside the application source code.
The original version of Lotto is a 238 LOC application. \var obfuscation enlarges it to 291 LOC.

The second application, named \emph{Lottery}, is a client-server lottery game similar to bingo that works as follows:
in order to extract the bingo numbers, the client contacts the server asking for a challenge;
the server generates a random sequence of bytes and sends it back to the client;
the client uses the challenge as a random seed to derive a sequence of seven numbers whose value ranges
between 1 and 39 - these are the seven extracted numbers.
To claim a win, the customer exhibits its number sequence.
The winning sequence is sent back to the server that checks its validity against the
previously sent challenge; if the sequence is valid, the server accepts it and prints it in a log -- so, the win can be delivered to the customer; otherwise, the server rejects the sequence and exits. This iteration is repeated ten times in the application to be attacked.

The original version of Lottery is composed of 62 client LOC  and 498 server LOC.
\var obfuscation makes the client  as large as 84 LOC and the server 521 LOC.
The client  uses a support library, for communication purposes only, consisting of 452 LOC. This library is not protected with \var.

The complexity of the two systems has been designed to be different.
In fact, despite the number of LOC,  Lottery  is harder to understand and to modify, because it involves server-side logic that cannot be inspected by the subjects.

Subjects are asked to carry out an \emph{attack task}.
For the Lotto application, the attack task is to determine the jackpot sequence, by spotting where the hardcoded array with the sequence is declared and defined, and reporting its values (it is unlikely that subjects are able to guess by random choice the winning sequence in the allotted time).
For the Lottery application, the attack task is to modify the application to force the client to only extract numbers between 1 and 20.
The attack tasks do not depend on clear and obfuscated application, they only depend on the application (Lotto or Lottery).

We overall identify the tasks as follows:


\begin{LaTeXdescription}

	\item[ {$\task[1]$} ]  the attack  ported on the \emph{clear Lotto} application;
	\item[ {$\task[2]$} ]  the attack  ported on the \emph{obfuscated Lotto} application;
	\item[ {$\task[3]$} ]  the attack  ported on the \emph{clear Lottery} application;
	\item[ {$\task[4]$} ]  the attack  ported on the \emph{obfuscated Lottery} application;

\end{LaTeXdescription}

The experiment is performed in a unique session divided into two sub-sessions. Each subject undergoes two
distinct tasks, one per sub-session.
The session lasted 3 hours and 30 minutes, the sub-sessions have approximately lasted 1 hour and 45 minutes.
The task assignments were fully balanced,  to avoid the influence of obfuscation and application across sub-sessions:
each subject never worked on neither the same application nor the same treatment (obfuscation) in the two subsequent tasks.
Therefore, during the second sub-session, each subject received the other application in clear if the first application was obfuscated ($\task[1]\leftrightarrow\task[4]$ and $\task[2]\leftrightarrow\task[3]$).
Table~\ref{tab:subseq-tasks} summarises the four combinations of  tasks assigned in the two sub-sessions.
In addition, we paid attention to assign each task to the same proportion of subjects in the first -- hence,
also in the second -- sub-session.


\begin{table}[]
	\caption{Task assignments}
	\label{tab:subseq-tasks}
	\centering
	\begin{tabular}{@{}rcc@{}}
		\toprule
		& First sub-session task & Second sub-session task \\ \midrule
Group 1	&		$\task[1]$             & $\task[4]$              \\
Group 2	&		$\task[2]$             & $\task[3]$              \\
Group 3	&		$\task[3]$             & $\task[2]$              \\
Group 4	&		$\task[4]$             & $\task[1]$              \\ \bottomrule
	\end{tabular}
\end{table}

\subsection{Variables}
This section describes the variables used to perform the evaluation of the experiment.
As \emph{dependent variables}, we consider the following aspects of the executed attack tasks:

\begin{LaTeXdescription}
	
	\item[ \emph{Correctness}] of a performed attack task.
	The correctness variable is evaluated as:
    $$
		\mathit{Corr{(\task, s_i)}} =
			\begin{cases}
				1 & \text{if subject } s_i \text{ succeeded in task } \task\\
				0 & \text{if subject } s_i \text{ failed in task }  \task \\
			\end{cases}
	$$


	\item[ \emph{Time}] to perform an attack task.
        The variable $Time{(\task, s_i)}$ is measured as the number of minutes spent by  subject $s_i$ to perform  task $\task$, successfully or not.

	
	\item[ \emph{Efficiency}] of an attack task. The efficiency variable, related to  task $\task$, is the sum of the inverses of the time for all those subjects
	who successfully performed the task $\task$.
	Formally:
    \[
		\mathit{Eff}_{\task} = \sum_{s_i\in S_{\task}} \frac{Corr{(\task,s_i)}}{Time{(\task,s_i)}}
	\]
	where $S_{\task}$ is the set of subjects that performed the task $\task$.
Note that, even if the sum ranges on all the subjects involved in  task $\task$, the numerator function $\mathit{Corr}$ excludes from the computation each subject $s_i$ who failed the task $\task$ (i.e., $\mathit{Corr{(\task,s_i)}} = 0$).
	%
	
\end{LaTeXdescription}

We observe that the efficiency considers only the successful attacks.
A measure of efficiency considering all cases would be equivalent to the inverse of \emph{Time}.
We analyzed that variable too but decided not to report it because it brings no additional insights.
Note that all the dependent variables are related to the duration of the experiment, i.e., 	the quantity they measure is not independent from the time we have assigned to students for the tasks.
That is, the more the time of the task, the more the students that can correctly complete it, and the higher the average time and the efficiency.
This factor has been considered during the design and does not affect the results of our study, which aims at assessing the effectiveness of \var obfuscation by computing dependent variable on clear vs.\ obfuscated programs in equivalent conditions.

As \emph{independent variables}, we consider the following ones:
\begin{LaTeXdescription}

	\item[\emph{Treatment}] applied to the source code,
	i.e., whether  obfuscation was applied to the code or not. This is the main factor
	in our design.
	
	\item[\emph{Application}] used in a task; this can be used to understand
	how code complexity influences the time or the correctness of the attack task.
	
	\item[\emph{Lab}] is the order of the sub-sessions in the experiment; this is required to
	assess the learning across subsequent tasks, in terms of how the experience
	gained during the first assigned attack task influences the behaviour observed in the second task.
	
	\item[\emph{Experience}] of the subjects, in terms of number of years they	
	have practiced C language programming.

\end{LaTeXdescription}

\subsection{Hypotheses}
We can formulate the following null hypotheses to be tested:

\begin{itemize}
		
	\item $H_{01}$:  \var source code data obfuscation has no effect on the
		correctness of an attack.

	\item $H_{02}$:  \var source code data obfuscation  has no effect on the
		time to perform an attack.

	\item $H_{03}$:  \var source code data obfuscation has no effect on the
		efficiency of an attack.

\end{itemize}

\subsection{Materials and procedures}

In this section we detail the procedure followed and the material used during the experiments.
To perform the tasks, subjects have been provided with a PC equipped with the \cb IDE running on Windows 7.
We selected \cb because all the subjects were familiar with it, having used it in different courses during their Bachelor and Master Degrees. Before starting the experiment, the following materials were distributed to the participants:

\begin{itemize}
\item the experience questionnaire, used to acquire knowledge about the experience of the subjects in C programming and reverse engineering;
\item a description of the program to be attacked;
\item a zip archive containing a \cb project with the  program to be attacked. We provided a working and tested \cb project to prevent subjects from losing time in creating the project and avoid problems in the compilation and execution of the programs due to misconfigurations of the project.

For tasks involving the Lottery program, the executable file of the server was given, along with a batch file which automated its execution; the server has been executed locally, in order to avoid network related problems; we did not provide the server source code nor tools to tamper with binaries, since in a \mate scenario, the attacker does not have access to the server code of client-server applications;

\item the description of the attack task 
the subjects were asked to perform.
\end{itemize}

Before performing the assigned task,  participants had to fill in the experience questionnaire. The items in the questionnaire include:
\begin{enumerate}
\item the work experience as a professional programmer;
\item the overall experience in C programming, expressed in years;
\item how long participants have been using an IDE for C programming;
\item their experience in using a C debugger, in terms of actions they are able to perform with it:
\begin{itemize}
\item add breakpoints;
\item execute the program stepwise;
\item inspect the call stack;
\item inspect the program variables;
\end{itemize}
\end{enumerate}

After the questionnaire,  subjects were asked to execute the given task, following the procedure below:
\begin{enumerate}
\item read a brief introduction that describes the program to ne attacked;
\item install the \cb IDE, using an automated procedure made available on all PCs provided to subjects;
\item download the zip file, containing the \cb project from a given URL;
\item extract the \cb project from the archive file;
\item open the project with the \cb IDE;
\item (\emph{for the Lottery program only}) start the batch file that runs the server;
\item build and execute the program (the client for the Lottery program);
\item (\emph{for the Lottery program only}) look at the log file produced, trying to interpret the logged information;
\item read the description of the assigned task;
\item write down the task start time;
\item execute the task;
\item provide evidence that the task has been correctly executed:
\begin{itemize}
\item for the Lotto program, write down the jackpot sequence;
\item for the Lottery program, show  the assistants running the experiments the contents of the extractions accepted and logged by the server. As explained in Section~\ref{sec:objects}, the server stores in this file all  legal extractions received from the client, therefore  assistants checked that the file contains only extractions with numbers between 1 and 20, so as to make sure that the subject has successfully completed the task;
\end{itemize}
\item write down the task completion time; we asked the students to not take 
into account the time elapsed for the initial setting of the experiments, and 
for eventual questions asked to the assistants; they therefore reported an 
estimation of the actual time spent solving the tasks. 
\end{enumerate}

Finally,  participants had to fill in a post-experiment questionnaire, asking for their impressions on the task just completed.
The questionnaire includes the following items:
\begin{enumerate}
\item whether the task was clear to the subject;
\item whether there was enough time to perform the task;
\item whether the subject felt that the task was easy to perform;
\item which tools the subject used to perform the task:
\begin{itemize}
\item disassembler;
\item IDE;
\item debugger;
\item internet search;
\item other (open for the subject to specify);
\end{itemize}
\item on which activity most of the time was spent:
\begin{itemize}
\item reading and understanding the binary;
\item inspecting the execution by means of the debugger;
\item changing the execution by means of the debugger.
\end{itemize}
\end{enumerate}
The first three items were measured on a Likert scale with 5 levels.

\subsection{Analysis method}
\label{sec:analysis}
The experimental measures are first summarised with  basic descriptive statistics.
Correctness is reported in terms of proportion of correct answers.
For Time and Efficiency, we report the mean and standard deviation.

In all hypothesis testing we  consider as independent variables the main factor -- i.e.\ Treatment -- and two co-factors, Application and Lab.

Among the two co-factors, Application is potentially  the most relevant: in fact, we are considering two applications whose size (and possibly complexity) is quite  diverse: a 238 LOC stand-alone vs.\ 62 LOC client-server application (in the Clear version). For this reason, we will report it in the summary tables and diagrams together with the main factor.

The Lab co-factor, representing the order of the lab task might be relevant in case maturation or learning effects emerge.

The two potentially confounding co-factors are included in all the hypothesis testing analyses. The motivation is that we aim at assessing the effect of the main factor once the co-factors have been accounted for.

To test hypothesis $H_{01}$, concerning  Correctness, we  use a logistic regression of Correctness vs. the three independent variables. Such analysis is suitable for the dichotomous nature of the measure.
The logistic regression is based on the following model:

$$
  Correctness = \frac{1}{1+e^{-(\beta_0 +
                              \beta_{T} \cdot T +
                              \beta_{A} \cdot A +
                              \beta_{L} \cdot L
                              )}}
$$
where $T$ and $A$ are indicator variables for the Treatment and Application variables, while $L$ indicates the assignment order. In particular:

$$
\begin{array}{l}
T = \begin{cases}
      1 & \text{if } Treatment = \text{Obfuscated} \\
      0 & \text{if } Treatment = \text{Clear} \\
    \end{cases}
\\
\\
A = \begin{cases}
      1 & \text{if } Treatment = \text{Lotto} \\
      0 & \text{if } Treatment = \text{Lottery} \\
    \end{cases}
\\
\\
L = \begin{cases}
      1 & \text{if first task} \\
      2 & \text{if second task}  \\
    \end{cases}
\end{array}
$$

To test hypotheses $H_{02}$ and $H_{03}$ concerning Time and Efficiency, we conduct a non-parametric test equivalent to ANOVA -- permutation test -- of the output variable vs. the three factors. The choice of a non-parametric test method is due to the expected non-normality of the measures.
The linear regression is based on the following model:

$$
  V = \beta_0 + \beta_{T} \cdot T +
            \beta_{A} \cdot A +
            \beta_{L} \cdot L
$$

Where $V$ is the output variable (either Time or Efficiency) and the other variables are the same as those used in the logistic regression.

The assessment of the statistical test results is carried out assuming significance at a 95\% confidence level ($\alpha$=0.05).
Since we test three distinct hypotheses on the same set of participants, to avoid inflating the family-wise error rate we applied the Bonferroni correction; therefore we  employ a corrected $\alpha_C=0.05/3=0.017$ for decisions.
So, we reject the null-hypotheses when $p\mathrm{-value}<\alpha_C$. 

All the data processing is performed with the R statistical package~\cite{R}. In particular the permautation test analysis was conducted using the lmPerm package \cite{lmPerm}.

%

\begin{table*}[t]
\centering
\caption{Summary statistics for Correctness and Time.}
\label{tab:descriptive}
\begin{tabular}{@{}llrrrrrrr@{}}
\toprule
  & & & \multicolumn{2}{c}{Correctness} & \multicolumn{2}{c}{Time}& \multicolumn{2}{c}{Efficiency} \\
\cmidrule(lr){4-5}\cmidrule(lr){6-7}\cmidrule(lr){8-9}
Treatment & Application & N & \quad n & prop. & mean & sd & mean & sd \tabularnewline
\midrule
Clear & Lottery & 7 & 3 & 0.43 & 30.00 & 14.50 & 3.98 & 2.03  \tabularnewline
Clear & Lotto & 6 & 6 & 1.00 & 6.50 & 3.08 & 10.75 & 4.05  \tabularnewline
Obfuscated & Lottery & 7 & 1 & 0.14 & 94.57 & 14.37 & 0.65 & NA  \tabularnewline
Obfuscated & Lotto & 7 & 5 & 0.71 & 56.43 & 33.15 & 1.61 & 1.90  \tabularnewline
\bottomrule
\end{tabular}
\end{table*}

\section{Results}\label{results}

Before starting the analysis we looked for potential outliers. One subject qualified as such, showing an outlier timing (in excess) for the easiest task and an outlier timing (in defect) for the hardest task.
This is probably due to a mistake in the annotation of the time.
We decided to discard the subject from any further analysis.

Table \ref{tab:descriptive} reports the descriptive statistics for the three output variables for different combination of Treatment and Application levels.

\subsection{$H_{01}$: Correctness}\label{correctness}

\begin{figure*}[tb]
\centering
\subfloat[Proportion of correctly completed tasks per Treatment and Application.]{
\includegraphics[width=0.3\textwidth]{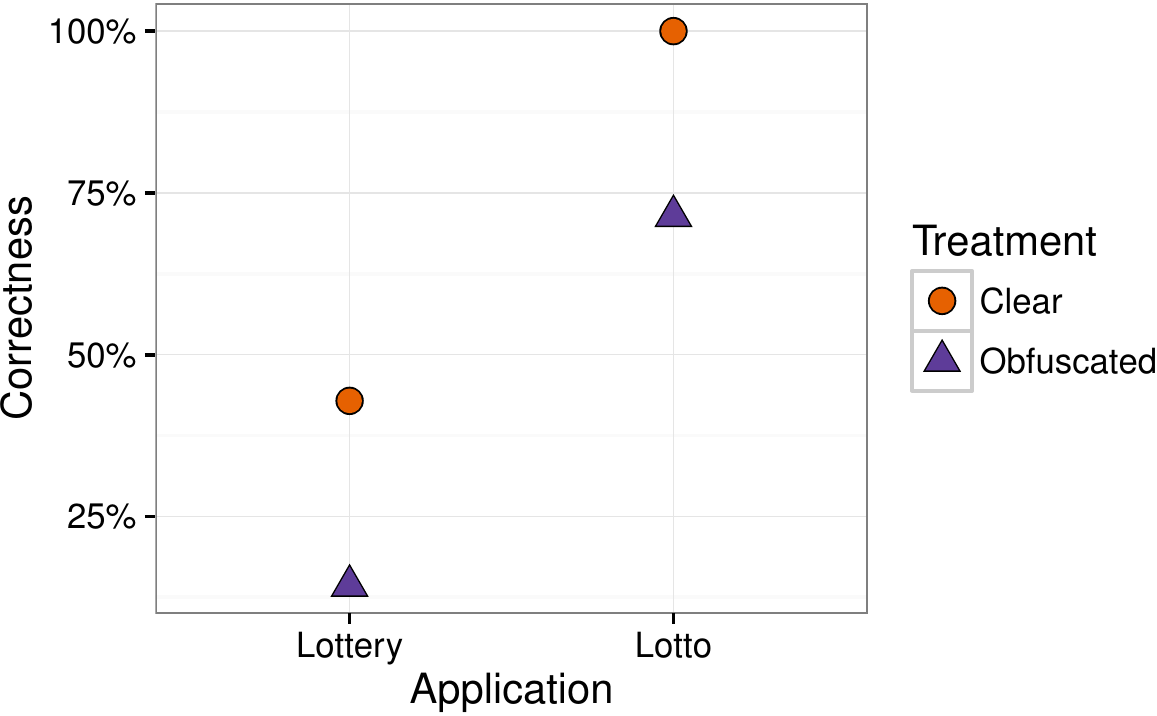}
 \label{fig:PlotCorrectness}}\quad
\subfloat[Boxplot of time to complete the task.]{
\includegraphics[width=0.3\textwidth]{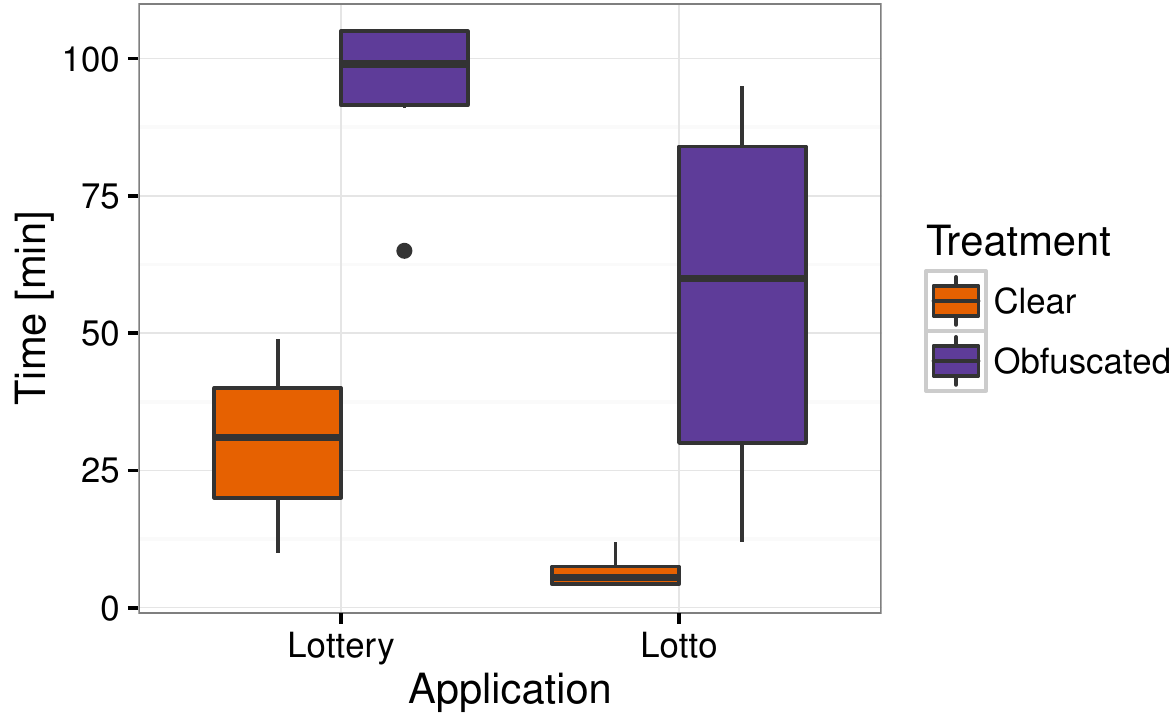}
 \label{fig:PlotTime}}\quad
\subfloat[Efficiency of attacks.]{
\includegraphics[width=0.257\textwidth]{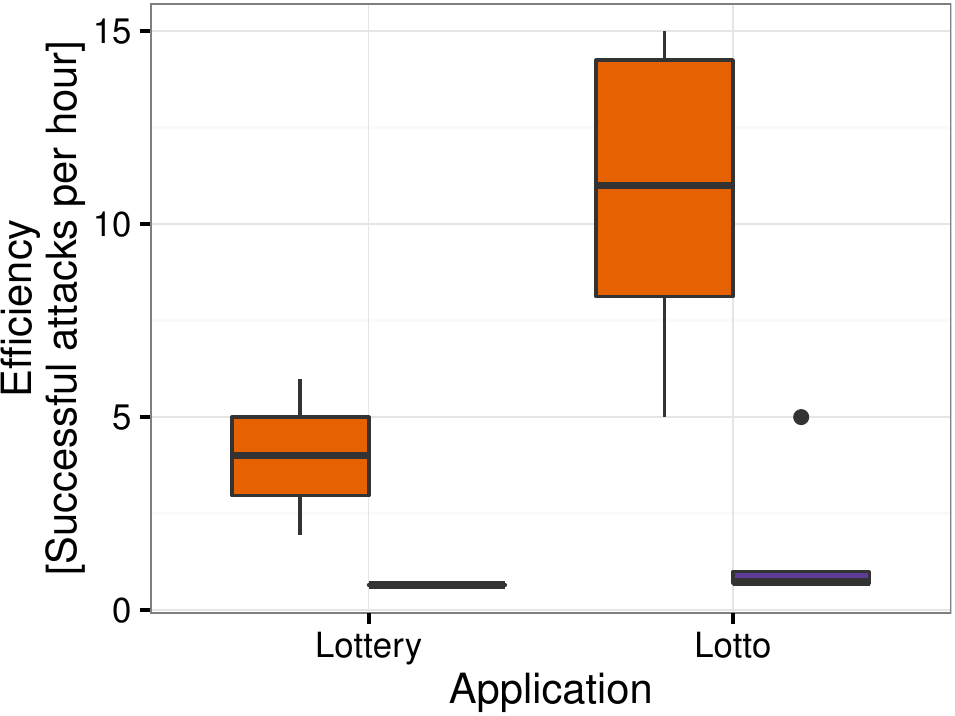}
 \label{fig:PlotEff}}
\caption{Outcomes of the experiment}
\end{figure*}

The effect of the treatment on the Correctness is visible in the dot-plot reported in Figure \ref{fig:PlotCorrectness}.
The results of the tests on the logistic regression are reported in Table \ref{tab:Correctness}.
Based on the results from the tests, we cannot reject the null hypothesis $H_01$: the obfuscation Treatment has no significant effect on the Correctness of the attack task outcome.
Only Application significantly affects the correctness of the task.
The potential confounding factor Lab has no relevant effect.

\begin{table}
\centering
\begin{small}
\caption{Logistic regression of Correctness}\label{tab:Correctness}
\begin{tabular}{@{}lrrrr@{}}
\toprule
& Est. & Std.Err. & z value &
Pr(\textgreater{}\textbar{}z\textbar{})\tabularnewline
\midrule
$\beta_0$ (Intercept) & -0.101 & 1.991 & -0.051 & 0.960\tabularnewline
$\beta_T$ TreatmentObfuscated & -2.001 & 1.240 & -1.614 & 0.107\tabularnewline
$\beta_A$ ApplicationLotto & 3.235 & 1.246 & 2.595 & \textbf{0.009}\tabularnewline
$\beta_L$ Lab & -0.022 & 1.086 & -0.020 & 0.984\tabularnewline
\bottomrule
\end{tabular}
\end{small}
\end{table}

\subsection{$H_{02}$: Time}\label{h2-time}
The distributions of Time for different combinations of Treatment and Application are reported in the boxplot of Figure \ref{fig:PlotTime}.
The result of the permutation tests on the linear regression is reported in Table \ref{tab:Time}.


\begin{table}[tb]
\centering
\caption{Permutation test of Time}\label{tab:Time}
\begin{tabular}{@{}lrrr@{}}
\toprule
& Estimate & Iter & Pr(Prob)\tabularnewline
\midrule
$\beta_0$ (Intercept) & 35.566 & 5000 & \textbf{\textless0.001}\tabularnewline
$\beta_A$ ApplicationLotto & -32.799 & 5000 & \textbf{0.002}\tabularnewline
$\beta_T$ TreatmentObfuscated & 55.151 & 5000 & \textbf{\textless0.001}\tabularnewline
$\beta_L$ Lab & -9.511 & 316 & 0.241\tabularnewline
\bottomrule
\end{tabular}
\end{table}

We reject the null hypothesis $H_02$: the obfuscation Treatment has a
significant effect on the attack task time.

Similarly to what we observed for Correctness, also  Time is affected by the
specific Application considered in the task.

The potential confounding factor Lab has no significant effect.

The goodness of fit for the linear regression ( \(R^2=0.70\) ) is relatively high, considering the inherent difference among the participants.

\subsection{$H_{03}$: Efficiency}\label{h3-eff}

The distributions of Time for different combination of Treatment and Application are reported in the boxplot of Figure \ref{fig:PlotEff}.
The result of the permutation tests on the linear regression is reported in Table \ref{tab:Eff}.


\begin{table}
\centering
\caption{Permutation test of Efficiency}\label{tab:Eff}
\begin{tabular}{@{}lrrr@{}}
\toprule
& Estimate & Iter & Pr(Prob)\tabularnewline
\midrule
$\beta_0$ (Intercept) & 4.786 & 5000 & \textbf{0.006}\tabularnewline
$\beta_A$ ApplicationLotto & 5.424 & 1233 & 0.109\tabularnewline
$\beta_T$ TreatmentObfuscated & -7.717 & 5000 & \textbf{0.002}\tabularnewline
$\beta_L$ Lab & 1.025 & 112 & 0.477\tabularnewline
\bottomrule
\end{tabular}
\end{table}

On the basis of the tests, we can reject the null hypothesis $H_03$: the obfuscation Treatment has a significant effect on the attack Efficiency.
Neither Application nor Lab have any effect on the Efficiency.

The goodness of fit for the linear regression ( \(R^2=0.59\) ) is relatively high, considering the inherent difference among the participants.

A practical measure of how much Obfuscation is effective in reducing the efficiency of an attacker can be computed by dividing the mean Efficiency on the clear code by that on obfuscated code, for either applications.
The Efficiency ratios for the two applications are 6.1 for Lottery and 6.6 on Lotto.

\subsection{Post-questionnaire}\label{post-questionnaire}

\begin{figure*}[t]
\centering
\includegraphics[width=0.75\linewidth]{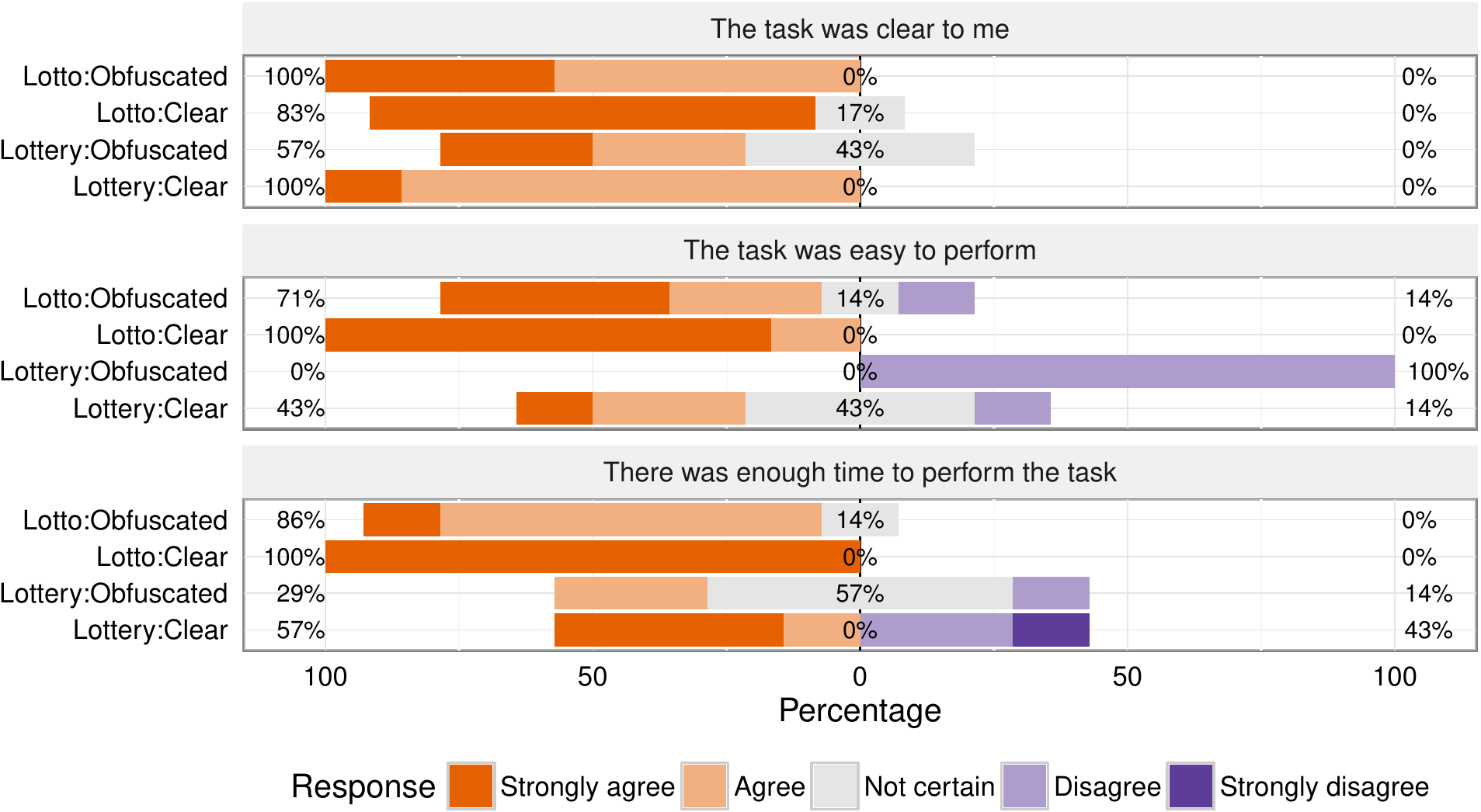}
 \caption{Assessment of task difficulties}\label{fig:PostDifficulties}
\end{figure*}

The post-questionnaire contains three items aimed at evaluating the \perceived difficulties encountered by the participants while performing the attack tasks.
The participant perception of clarity, availability of time, and easiness of the tasks is reported in Figure \ref{fig:PostDifficulties}.

We observe that in general the task assignments were considered clear;
somewhat less for the obfuscated Lottery.

Concerning the available time, when the participants worked on Lottery in a few cases they felt that not enough time was allowed. No problem of time was reported when they worked on Lotto.

The different complexity of the two applications shows up in the responses to the third item concerning easiness of the task. The attack task on Lottery was considered more difficult. In addition, as expected, the tasks on obfuscated code turned out to be more difficult than those on clear code.

\begin{figure*}[tb]
\centering
\subfloat[Tools used during the task.]{
\includegraphics[width=0.45\linewidth]{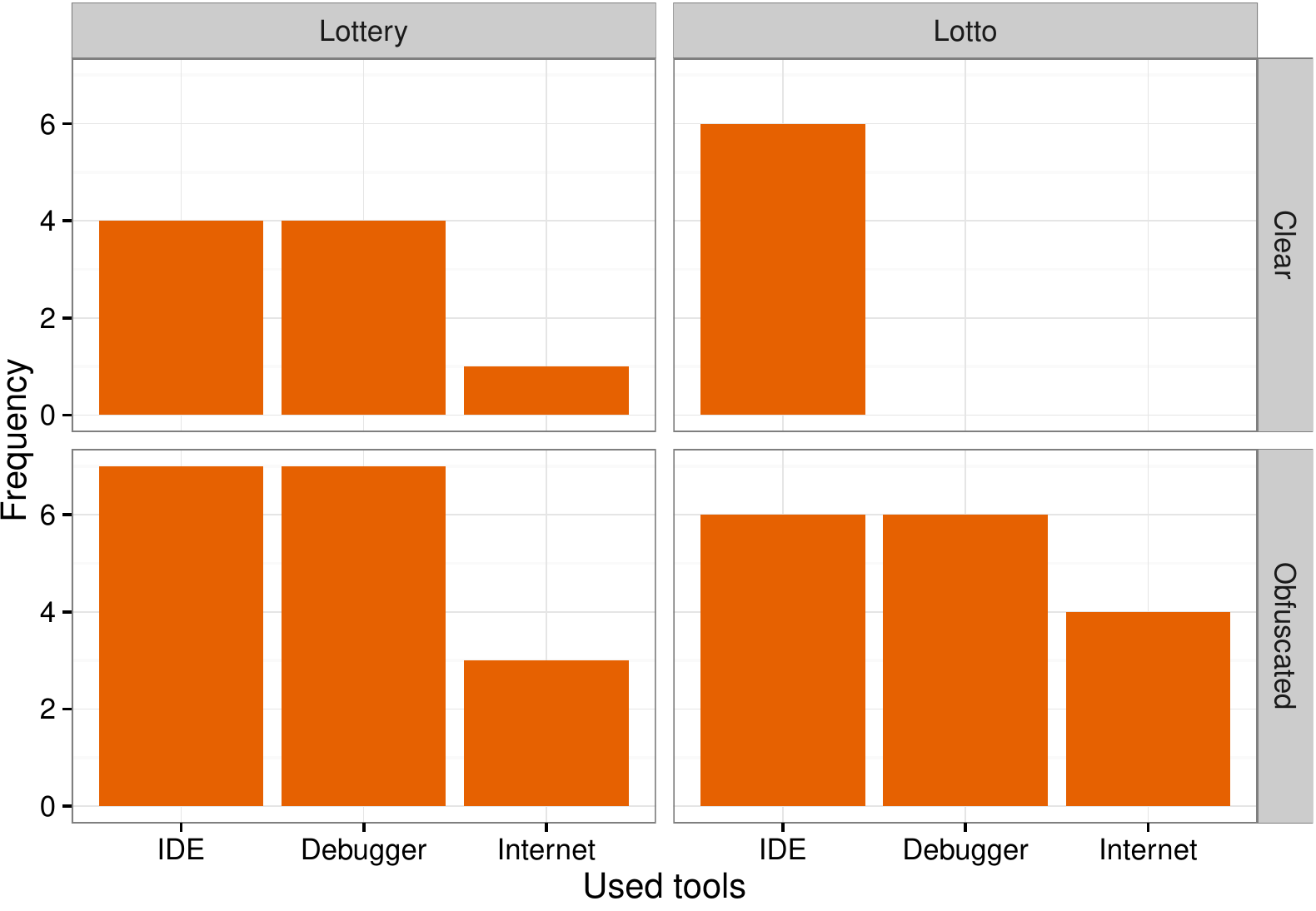}
 \label{fig:UsedTools}}\qquad
\subfloat[Activity that required most time.]{
\includegraphics[width=0.45\linewidth]{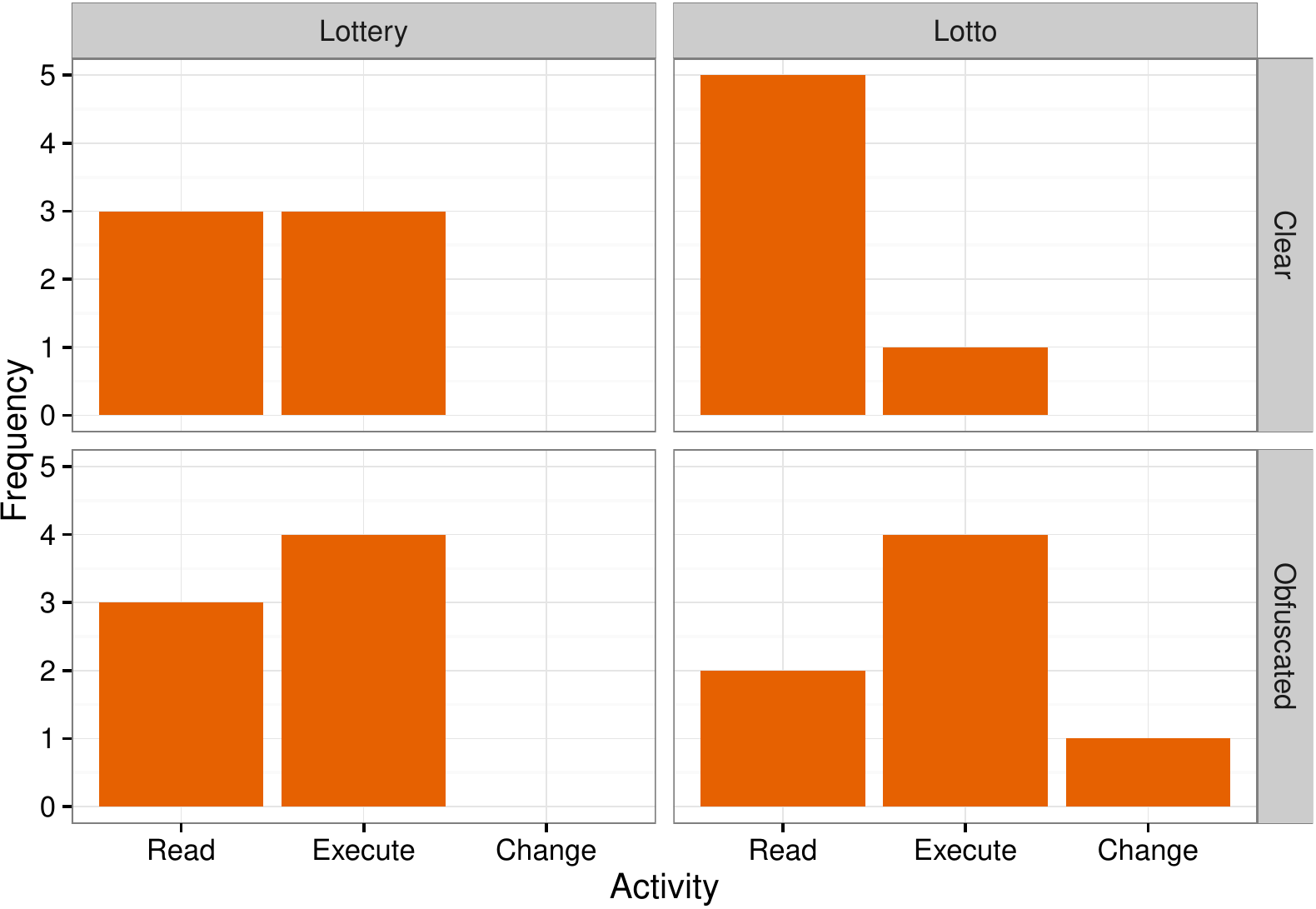}
 \label{fig:MostTimeActivity}}
\caption{Tools and activities analysis.}
\end{figure*}

The second part of the post-questionnaire concerns the tools used to perform the task. Figure \ref{fig:UsedTools} shows the frequency of usage of each tool, by Application and Treatment.

We can observe two changes that occurred when the obfuscated version was used in the task: first, tools were used more; second, specifically the usage of the debugger increased.
In addition, we observe different pattern of usage between the two applications. This reflects a difference that also emerged in the previous analyses.

Eventually, an item of the post-questionnaire addressed the activity on which  participants spent most of their time during the task (see Figure \ref{fig:MostTimeActivity}).

We can observe that the obfuscated versions of both applications required the participants to devote more time to execution of the application. This phenomenon appears in agreement with the increased usage of the debugger.
Moreover, not surprisingly, we see a difference between the applications.


\section{Discussion}

\label{sec:discussion}
In the next sections we comment on the results and their validity.

\subsection{Interpretation of results}

The analysis results are summarized below.
\begin{itemize}
\item Data obfuscation significantly affects the time to complete an attack task
(see Figure~\ref{fig:PlotTime})
and the attack efficiency (see Figure~\ref{fig:PlotEff}), while it does not affect the correctness of the attack outcome;
\item The two applications significantly differ both in terms of time and correctness, while no difference can be observed for the efficiency.
Such a disparity can be observed also in the \perceived difficulties of the task (see Figure~\ref{fig:PostDifficulties}).
\item The presence of obfuscation forces the attacker to resort more on all the available tools (see Figure~\ref{fig:UsedTools}).
\item Participants modified their attack strategy when facing obfuscated code: they mostly employed their time to execute the program. (see fig. \ref{fig:MostTimeActivity})
%
%
The data on the tools used (see fig. \ref{fig:UsedTools}) allows us to infer that they attempted to understand the behaviour of the program by observing its dynamic evolution. This is in line with the anecdotal knowledge about data obfuscation in general and \var in particular, which is expected to defeat attacks based on static analysis, while remaining relatively more vulnerable to dynamic analysis.
\item By looking at the mean efficiency reported in Table~\ref{tab:descriptive}, we can observe a six-fold decrease in attack efficiency when the \var obfuscation is used. Such effect is nearly the same for both applications.
\end{itemize}

We also observe that it is not possible to distinguish between subjects that did not succeeded in mounting the attack because they run out of time and the ones wouldn't have been able to actually perform the task. We highlight, however, that this distinction is irrelevant for our purpose of measuring the effect of \var.
Delaying successful attacks to the extent that they are no longer profitable is indeed one of the purposes of obfuscation techniques, which are not provable secure.

\subsection{Threats to validity}
\label{sec:threatstovalidity}


We have checked our experiments against the checklist of the possible threats to validity proposed by  Wohlin \etal \cite{wohlin00}, which are classified into construct, internal, conclusion, and external validity threats.

\textit{Construct validity} threats concern the relationship between the theoretical constructs and the actual metrics defined for the experiment.
While there are several types of attacks, we only focused on attacks that imply understanding and modifications of the source code. 
Alternative scenarios have not been tested.
The presence of obfuscation is expected to affect both the comprehension and the change activities.
However, in our experiment we could not measure the actual comprehension achieved by the subjects. Hence, we can only claim they achieved the minimum comprehension needed to perform their attack tasks, although we know they had to use more frequently sophisticated tools to understand the code when protections were applied.
Time is one of the direct measures we collected; it is a coarse grained metric, including both comprehension and change. Though the two activities might be separate, a typical attack consists of a close interleaving of the two.
The correctness of the attack task is evaluated as a boolean outcome. Although this is a very simple and crude metric, it reflects a real-case scenario where the attacker either gets access to the protected resources or not.
    %
    Finally, all the threats related to mono-operation and method are excluded by design.
    %

\textit{Internal validity} is concerned with the capability to capture a cause-effect relation between the independent variables and the outcomes.
That is, all noise factors, which may indirectly affect the outcomes, should have been eliminated or measured (e.g., assessed as negligible).
%
    Before starting any activity, the tasks and attack objectives have been explained to all subjects. The post-questionnaires confirmed that they had no problems in understanding the experiments.
A maturation effect during the experimental session could have occurred: every subject has been assigned two tasks in sequence. Although we did not report it for the sake of readability, we tested the results for statistically significant effects of the order of the task. No significant effect was found.
Since  subjects were very homogeneous we could divide them randomly into groups.
%
Finally, the experiment has been designed to avoid any ambiguity about direction of casual inference. Moreover, the inference stating that obfuscation renders attacks more complex did not appear ambiguous (and was already proved for analogous techniques in previous works).

\textit{External validity} threats are related to the impossibility to generalise our results to the case of real attackers who want to tamper with real applications protected by means of \var.
    The selection of participants was made on a voluntary basis: greater motivation better represents realistic hackers' profile.
    While students expertise in hacking programs is far from that of ''professional`` hackers, the problem solving ability of the best students is not supposed to be very different from that of hackers. Given a fixed time frame, the expertise certainly affects the correctness variable in tampering with a given application, but the selection of the applications to protect has been fine tuned to have enough successful attacks in the 105 minutes available for the experiments even with students.
    Furthermore, we measure our outcomes and draw conclusions on the effectiveness of \var by comparing the performance of subjects on clear and obfuscated versions of the application. Students with homogeneous expertise give the same validity results as hackers with homogeneous expertise. This comparison mitigates the lack of expertise.
    Our experiment included two applications. However, we do not have enough findings to estimate how \var may protect programs that are considerably different (e.g., larger or more complex) than the considered ones. This is one of the main directions for future work: including complexity metrics of the applications to protect as independent variables. 
    However, complexity metrics have been not considered in this experiment as, given the size of our samples, we wouldn't have drawn any significant conclusion.
    Tools made available to subjects were up to date and valid representative of tools hackers may use.
    When we ran the experiment, no special event or news could have affected the data collected from subjects. Actually, we were not acquiring any subjective data, but their expertise and impression on the attack tasks.
    We are aware that the effectiveness variable we have introduced is not the only way to measure the impact of successful attacks. Indeed, our formula favours attacks that are mounted quickly even if by a limited number of participants. We have decided for this approach as an application becomes vulnerable as soon as the the first attacker succeeds. We have found less interesting alternative formulas that emphasize when more people succeeds in mounting an attack with higher average time. Moreover, it did not add any insight to what we already presented in Section~\ref{sec:analysis}.


\textit{Conclusion validity} threats are related to the validity of the methods to derive outcomes from the treatment data.
We have used non-parametric statistical methods and controlled the error rate (permutation test ANOVA, error rate corrected with the Bonferroni method) as presented in Section~\ref{sec:analysis}.
We have collected data by means of survey questionnaires designed according to standard methods and scales \cite{oppenheim92}. 
Tasks were similar and balanced (one clear code and one obfuscated application; only \var obfuscation used); subjects were not heterogeneous, as they were all master students, and experiments avoided random irrelevance.

\section{Conclusions}
\label{sec:conclusions}

This paper reported an experiment aimed at assessing a specific data obfuscation technique -- {\var} -- in terms of its capability to hinder and delay an attack.
The experiment involved 15 students from the master degree programme in Computer Engineering at Politecnico di Torino.
The experiment revealed no significant difference in terms of attack success rate between obfuscated and clear programs. This is mainly due to the size of the sample.
In contrast, a significant difference was observed in the time required to complete the attack task.
In addition, the results show that the presence of the \var obfuscation is able to reduce by six times the attack efficiency (measured as the number of successful attacks per unit of time).
This outcome provides a practical clue that can be used in designing software protections based on data obfuscation.
However, since it is the first experiment that addresses data obfuscation techniques, we cannot draw conclusions on the comparison with other techniques.
Moreover, we had no possibilities to apply other techniques as given the size of the sample. 

In addition to the presence of obfuscation, the attack efficiency and time appear to be affected by the size and complexity of the program under consideration. This additional factor did not interfere, in our results, with the effect of obfuscation.
Though this outcome is far from being conclusive due to the limited range in size and complexity that was investigated in our experiment.

As further work, we plan to:
\begin{enumerate}
\item test the \var technique to other programs, with a wider variability in terms of size and complexity;
\item apply the same technique to binary programs;
\item apply and compare other obfuscation techniques;
\item involve subjects with different attacker skills.
\end{enumerate}

Of course, considering all these independent variables and confounding factors needs proper preparation of the experiments and high number of participants, which we cannot reach in our institutions.
Evaluating the effect of several obfuscation techniques on different applications can only be achieved with the collaboration and sharing of the results among researchers in the software engineering field.
We have started investigating how to build such a community, prepare a planning of experiments and share data.

\section*{Acknowledgement}
The research leading to these results has received funding from the European Union Seventh Framework Programme (FP7/2007-2013) under grant agreement number 609734.

\bibliographystyle{abbrv}
\bibliography{refs}

\begin{thebibliography}{10}

\bibitem{mate-attacks}
A.~Akhunzada, M.~Sookhak, N.~B. Anuar, A.~Gani, E.~Ahmed, M.~Shiraz,
  S.~Furnell, A.~Hayat, and M.~K. Khan.
\newblock Man-at-the-end attacks: Analysis, taxonomy, human aspects, motivation
  and future directions.
\newblock {\em J. Network and Computer Applications}, 48:44--57, 2015.

\bibitem{anckaert2007obfuscation}
B.~Anckaert, M.~Madou, B.~De~Sutter, B.~De~Bus, K.~De~Bosschere, and
  B.~Preneel.
\newblock Program obfuscation: a quantitative approach.
\newblock In {\em Proc. ACM Workshop on Quality of protection}, pages 15--20,
  2007.

\bibitem{barak2001pop}
B.~Barak, O.~Goldreich, R.~Impagliazzo, S.~Rudich, A.~Sahai, S.~Vadhan, and
  K.~Yang.
\newblock {On the (im) possibility of obfuscating programs}.
\newblock {\em Lecture Notes in Computer Science}, 2139:19--23, 2001.

\bibitem{ceccato2014large}
M.~Ceccato, A.~Capiluppi, P.~Falcarin, and C.~Boldyreff.
\newblock A large study on the effect of code obfuscation on the quality of
  java code.
\newblock {\em Empirical Software Engineering}, pages 1--39, 2014.

\bibitem{CeccatoPFRTT14}
M.~Ceccato, M.~{Di Penta}, P.~Falcarin, F.~Ricca, M.~Torchiano, and P.~Tonella.
\newblock A family of experiments to assess the effectiveness and efficiency of
  source code obfuscation techniques.
\newblock {\em Empirical Software Engineering}, 19(4):1040--1074, 2014.

\bibitem{ceccato2009effectiveness}
M.~Ceccato, M.~{Di Penta}, J.~Nagra, P.~Falcarin, F.~Ricca, M.~Torchiano, and
  P.~Tonella.
\newblock The effectiveness of source code obfuscation: An experimental
  assessment.
\newblock In {\em IEEE 17th International Conference on Program Comprehension
  (ICPC)}, pages 178--187, may 2009.

\bibitem{Collberg2009Surreptitious}
C.~Collberg and J.~Nagra.
\newblock {\em Surreptitious Software: Obfuscation, Watermarking, and
  Tamperproofing for Software Protection}.
\newblock Addison-Wesley Professional, 1st edition, 2009.

\bibitem{collberg1997taxonomy}
C.~Collberg, C.~Thomborson, and D.~Low.
\newblock A taxonomy of obfuscating transformations.
\newblock Technical Report 148, Dept. of Computer Science, The Univ. of
  Auckland, 1997.

\bibitem{falcarin2011guest}
P.~Falcarin, C.~Collberg, M.~Atallah, and M.~Jakubowski.
\newblock Guest editors' introduction: Software protection.
\newblock {\em Software, IEEE}, 28(2):24--27, 2011.

\bibitem{feigenspan2012measuring}
J.~Feigenspan, C.~K{\"a}stner, J.~Liebig, S.~Ape~l, and S.~Hanenberg.
\newblock Measuring programming experience.
\newblock In {\em Program Comprehension (ICPC), 2012 IEEE 20th International
  Conference on}, pages 73--82. IEEE, 2012.

\bibitem{goto2000quantitative}
H.~Goto, M.~Mambo, K.~Matsumura, and H.~Shizuya.
\newblock An approach to the objective and quantitative evaluation of
  tamper-resistant software.
\newblock In {\em Third Int. Workshop on Information Security}, pages 82--96.
  Springer, 2000.

\bibitem{linn2003obfuscation}
C.~Linn and S.~Debray.
\newblock Obfuscation of executable code to improve resistance to static
  disassembly.
\newblock In {\em Proc. ACM Conf.Computer and Communications Security}, pages
  290--299, 2003.

\bibitem{oppenheim92}
A.~N. Oppenheim.
\newblock {\em Questionnaire Design, Interviewing and Attitude Measurement}.
\newblock Pinter, London, 1992.

\bibitem{R}
{R Core Team}.
\newblock {\em R: A Language and Environment for Statistical Computing}.
\newblock R Foundation for Statistical Computing, Vienna, Austria, 2015.

\bibitem{sutherland2006empirical}
I.~Sutherland, G.~E. Kalb, A.~Blyth, and G.~Mulley.
\newblock An empirical examination of the reverse engineering process for
  binary files.
\newblock {\em Computers \& Security}, 25(3):221--228, 2006.

\bibitem{udupa2005deobfuscation}
S.~K. Udupa, S.~K. Debray, and M.~Madou.
\newblock Deobfuscation: Reverse engineering obfuscated code.
\newblock In {\em Proceedings of the 12th Working Conference on Reverse
  Engineering}, pages 45--54, Washington, DC, USA, 2005. IEEE Computer Society.

\bibitem{visaggio2013empirical}
C.~A. Visaggio, G.~A. Pagin, and G.~Canfora.
\newblock An empirical study of metric-based methods to detect obfuscated code.
\newblock {\em International Journal of Security \& Its Applications}, 7(2),
  2013.

\bibitem{lmPerm}
B.~Wheeler.
\newblock {\em lmPerm: Permutation tests for linear models}.
\newblock R package version 2.0.

\bibitem{wohlin00}
C.~Wohlin, P.~Runeson, M.~H\"ost, M.~Ohlsson, B.~Regnell, and A.~Wessl\'en.
\newblock {\em Experimentation in Software Engineering - An Introduction}.
\newblock Kluwer Academic Publishers, 2000.

\end{thebibliography}

\end{document}